# Objective dyspnea evaluation on COVID-19 patients learning from exertion-induced dyspnea scores

Zijing Zhang*, Jianlin Zhou, Thomas B. Conroy, Samuel Chung, Justin Choi, Patrick Chau, Daniel B. Green, Ana C. Krieger and Edwin C. Kan, *Senior Member, IEEE*

*Abstract—* **Objective: Dyspnea is one of the most common symptoms for many pulmonary diseases including COVID-19. Clinical assessment of dyspnea is mainly performed by subjective self-report, which has limited accuracy and is challenging for continuous monitoring. The objective of this research study is to determine if dyspnea progression in COVID patients can be assessed using a non-invasive wearable sensor and if the findings are comparable to a learning model of physiologically induced dyspnea on healthy subjects. *Methods:* Non-invasive wearable respiratory sensors were employed to retrieve continuous respiratory characteristics with user comfort and convenience. Overnight (~16h) respiratory waveforms were collected on 12 COVID-19 patients, and a benchmark on 13 healthy subjects with exertion-induced dyspnea were also performed for blind comparison. The learning model was built from the respiratory features with self report on 32 healthy subjects under exertion and airway blockage. *Results:* High similarity between dyspnea on COVID patients and physiologically induced dyspnea on healthy subjects was established. COVID patients have consistently high objective dyspnea scores in comparison with normal breathing of healthy subjects. We also exhibited continuous dyspnea scoring capability for 12-16 hours on patients. *Conclusion:* This paper validates the viability to use our objective dyspnea scoring for clinical dyspnea assessment on COVID patients. *Significance:* The proposed system can help the identification of dyspneic exacerbation in conditions such as COVID, leading to early intervention and possibly improving their outcome. This approach can be potentially applied to other pulmonary disorders such as asthma, emphysema, and pneumonia.**

## I. Introduction

Dyspnea, also known as the patient's feeling of difficult or labored breathing, is a clinical symptom nearly as important as pain, affecting a quarter of the general population and half of seriously ill patients [1][2]. Dyspnea can be a prevalent manifestation in conditions such as chronic obstructive pulmonary diseases (COPD), bronchitis, asthma, COVID-19, pneumonia, heart failure, and panic disorders [3]. Dyspnea can be further divided into acute onset and chronic dyspnea: the latter, by definition, has been present for more than four weeks. COVID-19 caused by severe acute respiratory syndrome coronavirus-2 (SARS-CoV-2) has rapidly spread across the globe since 2020. Over 30% of patients with COVID have experienced chronic dyspnea [4]. Dyspnea typically sets in between the 4[th] and 8[th] day of illness. Timing of dyspnea may be one of the most important hallmarks of more significant COVID infection, especially for clinicians seeing patients in an ambulatory setting [1]. Studies found that dyspnea, rather than fever [8], was significantly associated with higher mortality in COVID patients [6][7]. The initial days after the onset of dyspnea are critical for identifying progressively worsening conditions [1]. In addition to the dyspnea experienced by many patients during the acute phase of COVID infection, dyspnea may also be found in association with the longer-term sequelae post-COVID, or so called long COVID, which is thought to affect 10–50% of COVID survivors [9]. Dyspnea is also a frequent symptom of post-COVID syndrome (PCS) [10][11]. Some patients with persistent dyspnea after recovering from COVID also have documented decrease and/or dysfunction in myocardial performance [12].

Nevertheless, in present clinical practices, dyspnea is mainly assessed by self-reports from patients. Subjective dyspnea can be assessed in person or remotely [14] using patient interview, and augmented by surrogate measures such as the Medical Research Council (MRC) Dyspnea Scale [15] and Borg Scale [16]. Studies indicated that subjective dyspnea measures have inadequate accuracy in high-risk COVID patients, not only because the sensation is gradual and varied with time, but also the patients can become nervous after encountering positive test results [17], both of which can contribute to biases in the self-report. The subjective dyspnea score can also vary for each person based on emotion and tolerance, and can be challenging to assess for those who refuse to cooperate or cannot communicate due to medical issues such as stroke, dementia, and sudden loss of speech. Frequent queries to patients for continuous dyspnea evaluation are tedious and stressful, and essentially impractical for timely prognosis and diagnosis.

Shortness of breath highly correlates with pulmonary functions [18]. Surrogate measures of respiratory function can help determine dyspnea severity, however existing techniques have limitations. Pulmonary function tests can only capture respiratory measures at discrete point in time [19], and require patients' adequate effort and cooperation. Chest Computed Tomography (CT) can provide high-resolution images of the lung, however it is expensive, requiring specialized equipment that is not always available, providing only discrete lung

This work was supported by NSF RAPID 2033838.

Z. Zhang, J. Zhou, T. B. Conroy, and E. C. Kan are with School of Electrical and Computer Engineering, Cornell University, Ithaca, NY 14853, USA. S. Chung, J. Choi, P. Chau, D. B. Green and A. C. Krieger are with Center for Sleep Medicine at Weill Cornell Medicine, New York, NY 10065, USA. E-mails: {zz587; jz899; tbc38; eck5}@cornell.edu; { sjc9009; juc9107; pac2031; dag2017; ack2003 }@med.cornell.edu).

snapshots in a dedicated clinical setup, with limited utility for continuous assessment of lung function [20]. Respiratory inductance plethysmography (RIP) [21], strain gauge (SG) [22], and spirometer [23], can measure lung function however often can be uncomfortable, requiring connection to immobile machines, operator assistance, and patient cooperation. Thus they are not feasible tests for clinical settings, especially in conditions that require repetitive or continuous monitoring.

Considering the limitations of the current methods to monitor respiratory function and assess dyspnea, we propose to augment our previous study on the objective scoring of physiologically induced dyspnea [24], which used a non-invasive and wearable respiratory sensor and a machine learning (ML) model, to provide real-time objective dyspnea scores for COVID patients based on continuous respiratory metrics.

In this paper, we collected overnight clinical data of patients admitted to the hospital due to acute COVID ($N = 12$), using wearable respiratory sensors from the Weill Cornell Center for Sleep Medicine. These patients had confirmed pulmonary involvement based on radiological imaging. To benchmark the results, we also performed a separate human study ($N = 13$) on healthy participants using exactly the same experimental setup. We analyzed the statistic distribution of respiratory metrics from COVID patients and healthy controls, and demonstrated a high similarity between dyspnea on COVID patients and the physiologically induced dyspnea on healthy subjects. The features associated with breathing changes due to physical exertion were similar to those from pulmonary disorders. By training on our previous objective dyspnea scoring model ($N = 32$) on healthy subjects with induced dyspnea from exertion and airway blockage [24], we can further output continuous dyspnea scores of COVID patients using respiratory waveforms captured from wearable sensors to evaluate dyspnea severity for prognosis of dyspneic episodes, as well as rehabilitation after recovering from COVID. In the future, our respiratory sensor and objective dyspnea scoring system can be potentially applied to symptomatic evaluation of dyspnea in patients of asthma, pneumonia, and COPD.

## II. EXPERIMENTAL SETUP AND PROTOCOL

### A. Experimental setup

To monitor the respiration in hospitalized COVID patients with confirmed lung infection by chest imaging, we built an all-in-one wearable radio-frequency (RF) near-field sensor on a 4-layer printed circuit board (PCB), as shown in Fig. 1(a). The block diagram of the near-field RF sensing system is shown in Fig. 1(b). Two SimpleLink modules (Texas Instrument CC1310) with sub-1 GHz ultra-low-power wireless microcontrollers were used as the sensing transmitter (Tx) and receiver (Rx). The transceivers employed quadrature I/Q modulation, where two channels of 12-bit I and Q samples were sampled at 2 kHz and accumulated into one 400-bytes cyclic buffer at the sensor Rx. Once the I/Q buffer was filled, the Rx radio core would bundle the I and Q samples with the readings of temperature, accelerometer, and gyroscope to a micro secure digital (SD) card through the serial peripheral bus (SPI). The temperature sensor (Texas Instrument TMP112) and inertial measurement unit (Bosch Sensortec BMI160) were connected by the inter-integrated-circuits (I2C) serial protocol. The battery provided the system power through a low-dropout (LDO) module.

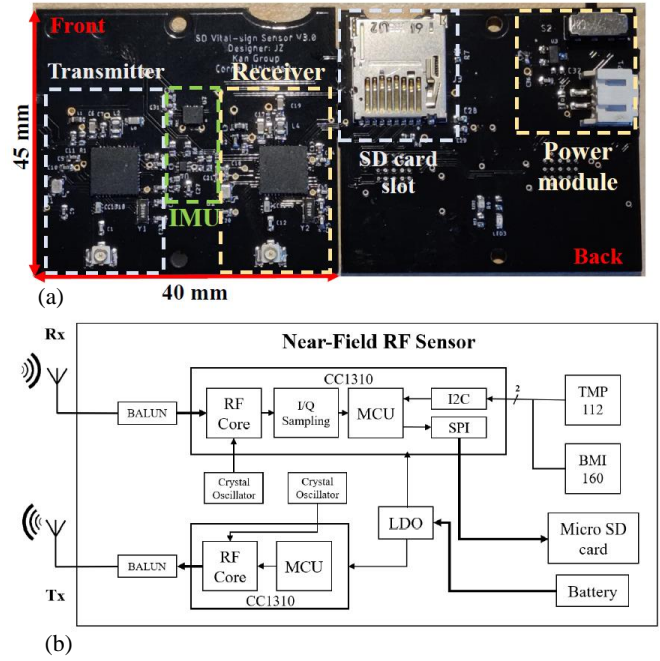

Fig. 1. The all-in-one wearable respiratory sensor on PCB: (a) The front and back photo; (b) The block diagram for the RF sensor.

### B. Data collection from COVID patients

Respiratory data acquisition from COVID patients took place at Weill Cornell Medicine July 2021 and March 2022. Patients admitted to New York Presbyterian Hospital with COVID-19 symptoms that had lung imaging studies were offered participation in the study. All participants signed an informed consent form. The study protocol was reviewed and approved by the Weill Medical Center Institutional Review Board (IRB Protocol #: 20-06022181).

Upon enrollment, patient demographic, health status, and baseline vital-sign data were gathered. Health status information included hypertension, obstructive sleep apnea, cancer history, asthma, COPD, and other chronic lung diseases. Baseline vital-sign data included heart rates, breathing rates (BR), temperature, and oximetry $SpO_2$. Other recorded information included medications administered or the presence of ventilation or supplemental oxygen during recording. After the admission information was gathered, the medical staff applied the ApneaLink device (Resmed ®) with two NCS sensing units on the patient's chest for overnight monitoring. The two sensors were synchronously powered on to begin recording. Patients wore the sensor for an average of 14.3 hours overnight. The demographic distribution of COVID patients is shown in Supplementary Table I.

The setup could be removed by the medical staff at any point to allow for appropriate medical care. This active medical environment created uncertainty in sensor positioning as not all sensor removals were recorded and thus information was based on the recording from the sensor data. Additionally, chest belt tension and positioning were not recorded upon each mounting and removal, so the exact sensor conditions cannot be known. However, the uncertainty in an active clinical environment also implies the resulting data analyses can be representative of typical clinical or at-home use where strict sensor and behavior restrictions cannot be enforced, and therefore should enhance the generalizability of our findings.

*C. Healthy participant study protocol*

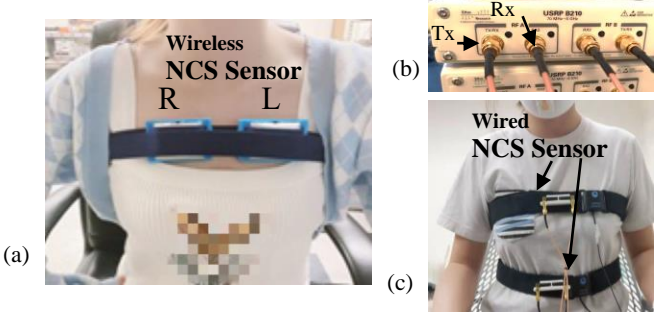

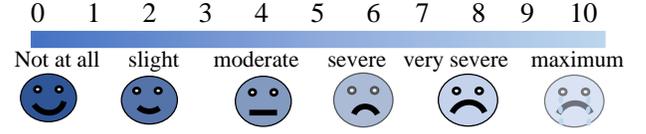

Fig. 2. Experimental setup: (a) Body deployment of two wireless NCS units by a chest belt; (b) SDR transceivers used for wired NCS sensors. (c) Wired wearable NCS sensors that were connected by cables to the sensing antennas on the chest and abdomen of a participant.

TABLE I. ACQUISITION OF DIFFERENT DATASETS

|   | Participants | Recording Time | Sensors |
|---|---|---|---|
| COVID | 12 COVID patients | Continuous 14 hours | Portable NCS sensors with accelerometers. |
| Exp. 1 | 13 healthy subjects | 1. Normal (30 mins) 2. Post-exercise (5 mins) | Portable NCS sensors with accelerometers. |
| Exp. 2 [24] | 32 healthy subjects | 1. Normal (5 mins) 2. Post-exercise (5 mins) | Wearable NCS software-defined radios. |

To further investigate the correlation between the dyspnea of COVID patients and physiologically-induced dyspnea on healthy subjects, we conducted another human study as indicated in Table I (Exp 1) on 13 healthy participants reporting dyspnea scores and measuring respiratory behaviors with dyspnea induced by exercise. For fair comparison without concerns on hardware difference, we used the same wearable respiratory sensors as in the COVID data collection. Fig. 2(a) shows the experimental setup with the participant wearing two sensors on left and right. The vertical position of two sensors is at the level of the sternum, roughly between the 3$^{rd}$ and 7$^{th}$ ribs. The demographic distribution of 13 healthy participants is shown in Supplementary Table II. The human study has been approved by Cornell Institutional Review Board (IRB) Protocol ID #1812008488. Written informed consent to take part in the study was obtained from all participants. Participants were instructed to follow a set of routines as listed in Table I. The participant first sat on a chair in a relaxed mode for normal breathing of 30 mins. To induce dyspnea, the participant would follow a 5-min cardio exercise video [25]. The participant would then sit back to the chair and be recorded for 5-min post-exercise breathing. The participants were asked to report subjective dyspnea scores $D_{self}$ several times in the Borg visual analog scale (VAS), as shown in Fig. 3 [26], during the transition points of the study –. The Borg scale is widely used in clinical assessement for dyspnea: 0 represents no dyspnea sensation at all, while 10 indicates maximum level of dyspnea.

Fig. 3. Description of the self-reported Borg visual analog scale (VAS) for dyspnea evaluation.

As shown in Table I and Figs. 2(b)(c) [24], we also adopted our previous dyspnea study for comparison, denoted as data from Exp 2. In this dyspnea study, we utilized the software-defined radio (SDR, Ettus B210) to the Tx/Rx antennas as shown in Fig. 2(b). Two wired NCS sensors were placed on the chest and the abdomen in the front torso. In this human study, participants first recorded 5-min normal breathing sitting on a chair, then used aerobic exercise to introduce dyspnea, and then recorded another 5 mins of post-exercise breathing. The dyspnea was also induced through facemask to change the lung elastance. The Borg dyspnea score was reported several times throughout the routine.

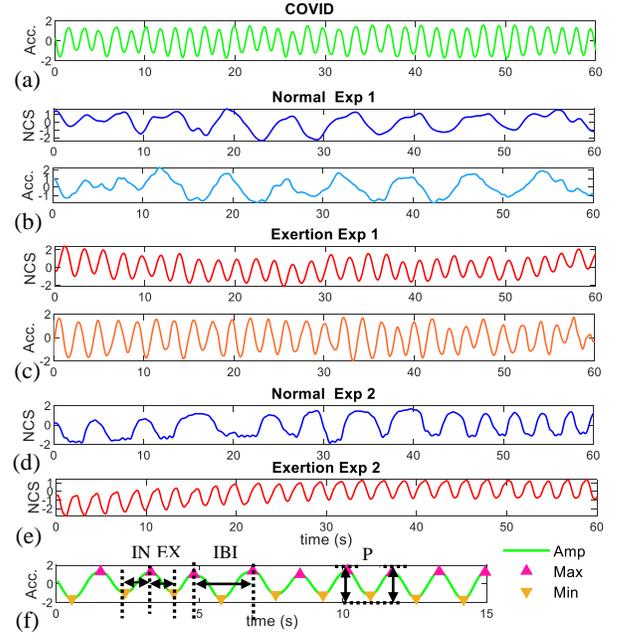

Fig. 4. Waveform examples: (a) COVID patients; (b) Healthy normal baseline breathing in Exp 1; (c) Healthy post-exertion breathing in Exp 1; (d) Healthy normal breathing in Exp 2; (e) Healthy post-exertion breathing in Exp 2; (f) Min-max peak detection for respiratory parameter extraction.

Fig. 4 presents examples of respiratory waveforms we acquired from different datasets. Y-axes are individually normalized in different channels. For the COVID dataset, we utilized the accelerometer channel for respiration monitoring. For Exp 1, we demonstrated NCS and accelerometer channels for both Normal and Exertion. For Exp 2, we demonstrated NCS recording for the same routines. Comparing normal breathing (b)(d) with exertion (c)(e) in Exp 1 and Exp 2, we can observe a consistent change in BR, with an increase in rate and a decrease in breath-to-breath variation post-exertion. We can also observe that COVID patients had higher BR than healthy participants during normal breathing, and the COVID waveforms were similar in frequency and amplitude to those acquired post-exertion in healthy participants. Respiratory parameters were extracted from the waveforms after min-max peak detection [24] as shown in Fig. 4(f), including inter-breathing intervals (IBI), inspiration intervals (IN), expiration intervals (EX) and peak-to-peak magnitude (PP).

## III. DATA PROCESSING

### A. Physiological analysis of dyspnea

The main purpose of respiration is to supply oxygen to body cells through the pulmonary circulation, with the auxiliary functions of producing sound, sniffing, and clearing of airway by coughing and sneezing. Respiration can be initiated involuntarily and voluntarily, and the voluntary part can be trained. When the blood oxygen saturation (SaO2) is low or $CO_2$ high, the breathing will be triggered to increase ventilation. However, when the body cannot respond properly due to various reasons such as airway obstruction, insufficient ambient oxygen, weakened respiratory muscles, cardiopulmonary disorders, or voluntary control for speaking, singing or breath holding, the sensation of dyspnea arises. To increase lung ventilation, often BR and lung volume (LV) will increase by panting or deep breathing. Alternatively, the inhalation and exhalation intervals will be adjusted depending on the muscle condition, airway obstruction, and ambient factors. As the respiratory reaction to dyspnea can be trained to reduce the uncomfortable feeling, similar to experiential avoidance for coping with pain, another common physiological reaction to dyspnea is the reduction of variability in successive breaths together with speaking restraint, when the body tries hard to use the best known breathing cycle to reduce the discomfort from dyspnea.

One major symptom for COVID patients is the dyspnea, which relies primarily on self-report sensation at the present practice. The dyspnea sensation is often derived from the decreased ventilation efficiency caused by pneumonia or related bronchitis. A distinct phenotype in long COVID is that patients have reduced exercise tolerance and experience exertional dyspnea more easily, even though major pulmonary parenchymal and airway abnormalities cannot be identified with chest imaging [27][28].

### B. Data preprocessing

After gathering overnight recording of COVID patients and performing comparison dyspnea study on healthy participants, we pre-processed our datasets, and then extracted the respiratory features to feed into the ML algorithms for dyspnea classification and scoring. We used MATLAB for signal processing and feature extraction, and python for ML algorithms.

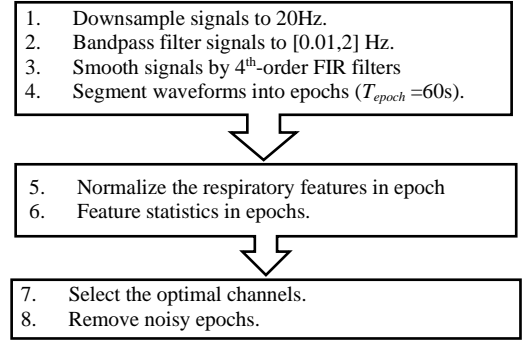

Fig. 5. Processing procedures of respiratory datasets in COVID patients and healthy participants.

For datasets in COVID and Exp 1 described in Table I, we retrieved respiratory waveforms from 1 NCS (amplitude) and 6 accelerometer (translational and rotational) channels. For Exp 2, we utilized multiple NCS channels from thorax and abdomen. Different channels and different datasets went through the same signal processing procedure for fair comparison. As shown in Fig. 5, we first down-sampled all datasets to 20 Hz, and then bandpass-filtered waveforms from 0.05 Hz to 2 Hz to remove the DC drift and high-frequency noises. Savitzky-Golay 4th-order finite impulse response (FIR) smoothing filter [29] was further employed to perform a local polynomial regression to smooth the waveform. For the long recording in each dataset, we opted to segment waveforms into short epochs of $T_{epoch} = 60$ s and a sliding window of $T_{slide} = 30$ s for feature extraction.

### C. Respiratory features

After epoch segmentation, waveforms were normalized to center at 0 with standard deviation of 1.0 in each epoch for every channel. Then we extracted features in each epoch for data analysis and constructed the ML model in the next section. We implemented the peak detection algorithm [30] by tracing a constantly updated moving-average curve in a given window. Then local maximum and minimum were accordingly labelled for parameter extraction. An example was shown in Fig. 4(f), where the green line was the filtered respiratory waveform from the COVID patient, and the red and yellow triangles marked the maximum and minimum peaks detected by the moving-average algorithm. The false peaks caused by the noise were mostly avoided. Respiratory parameters in each breath cycle were extracted to represent the instantaneous respiratory characteristics, as shown in Table II.

After gathering respiratory cycles and parameters, we extracted 37 respiratory features as shown in Table III. The first three features were: 1) mean ($\mu$); 2) standard deviation ($\sigma$); 3) coefficient of variation ($CoV$) of the respiratory parameters in Table II, where $CoV$ was defined as

$$CoV = \left(\frac{\sigma}{\mu}\right)^2 \qquad (1)$$

showing the extent of variability in relation to the mean.

To further capture variability between adjacent breaths, $\mathfrak{R}$ was the autocorrelation in a time lag of one respiratory cycle to measure the successive similarity of the given respiratory parameter. ς representing the successive difference was defined as the mean absolute difference between adjacent cycles. Additionally, *Skew* and *kurt* measured the tailedness and asymmetry of each respiratory cycle, and were averaged over all cycles within the epoch. *Cycle* denoted the total number of detected respiratory cycles in the epoch, and *entropy* denotes the total randomness or entropy of the waveform.

TABLE II. INSTANTANEOUS RESPIRATORY PARAMETERS (7)

| Extracted Parameters | Description |
|---|---|
| Breath Rate (*BR*) | Inverse of the interval between two neighboring minima. |
| Peak-to-Peak (*PP*) | Lung volume represented by signal difference in successive peaks. |
| Inhalation Interval (*IN*) | Time difference between one minimum and the following maximum. |
| Exhalation Interval (*EX*) | Time difference between one maximum and the following minimum. |
| Inter-Breath Interval (*IBI*) | Interval between two neighboring maxima. |
| In- Ex Ratio (*IER*) | Inhalation/Exhalation interval ratio. |
| In- Ex Volume Ratio (*IEPP*) | Inhalation/exhalation volume ratio. |

TABLE III. RESPIRATORY FEATURES (37)

| $\mu_{BR}$ | $\mu_{PP}$ | $\mu_{IN}$ | $\mu_{EX}$ | $\mu_{IBI}$ | $\mu_{IER}$ | $\mu_{IEPP}$ |
|---|---|---|---|---|---|---|
| $\sigma_{BR}$ | $\sigma_{PP}$ | $\sigma_{IN}$ | $\sigma_{EX}$ | $\sigma_{IBI}$ | $\sigma_{IER}$ | $\sigma_{IEPP}$ |
| $CoV_{BR}$ | $CoV_{PP}$ | $CoV_{IN}$ | $CoV_{EX}$ | $CoV_{IBI}$ | | |
| $\Re_{BR}$ | $\Re_{PP}$ | $\Re_{IN}$ | $\Re_{EX}$ | $\Re_{IBI}$ | $\Re_{IER}$ | $\Re_{IEPP}$ |
| $\varsigma_{BR}$ | $\varsigma_{PP}$ | $\varsigma_{IN}$ | $\varsigma_{EX}$ | $\varsigma_{IBI}$ | $\varsigma_{IER}$ | $\varsigma_{IEPP}$ |
| $\mu_{skew}$ | $\mu_{kurt}$ | entropy | cycle | | | |

TABLE IV. FREQUENCY FEATURES (14).

| $\eta_{f1}$ | $\eta_{f2}$ | $\eta_{f3}$ | $\eta_{f4}$ | $\eta_{f5}$ |
|---|---|---|---|---|
| $p_{f1}$ | $p_{f2}$ | $p_{f3}$ | $p_{f4}$ | $p_{f5}$ |
| $f_{BR}$ | $f_{HR}$ | $SNR_{BR}$ | $SNR_{HR}$ | |

Apart from 37 respiratory features extracted from the time domain, we added 14 features extracted from the frequency domain as shown in Table IV. $\eta_{fi}$ and $p_{fi}$ ($i = 1\sim5$) represented the power in specific bandwidth divided by the total power in all frequencies and time-averaged power density (dB/Hz), respectively. The five chosen bandwidths were $f_1 = (0, 0.4)$ Hz (mainly breathing frequency range); $f_2 = (0.4, 1)$ Hz; $f_3 = (1, 2)$ Hz; $f_4 = (f_{BR} - 0.15, f_{BR} + 0.15)$ Hz; $f_5 = (f_{HR} - 0.15, f_{HR} + 0.15)$ Hz. $f_{BR}$ and $f_{HR}$ were first estimated from the average BR and heart rate (HR) provided by hospital reports for every patient, and then further refined to be the local BR and HR in every epoch by finding the maximal energy in the possible frequency band. Signal-to-noise ratios (SNR) in BR and HR were denoted by $SNR_{BR}$ and $SNR_{HR}$ which were calculated by the maximal energy on the $f_{BR}$ and $f_{HR}$ divided by the estimated noise power.

*D. Channel and epoch selection*

After segmentation and feature extraction, we selected the optimal channel and epoch from the datasets according to the estimated signal quality. For the accelerometer, we had 6 channels consisting of X, Y, and Z translational (acc) and rotational (gyro) motions, as shown in Fig. 6 for an example. Feature extraction was performed on every channel, and the optimal channel was decided by the least variation of respiration parameters. We can observe from the waveforms that most channels can get similar BR =35, but the channels with smaller $\sigma_{BR}$ and $\sigma_{PP}$, such as 'gyro X' and 'gyro Y', have more stable respiratory waveforms. Therefore, we opted to use the optimal channel by the minimum mean of all covariation features in BR, PP, IN, EX and IBI. In Fig. 6, the optimal channel is 'gyro Y'.

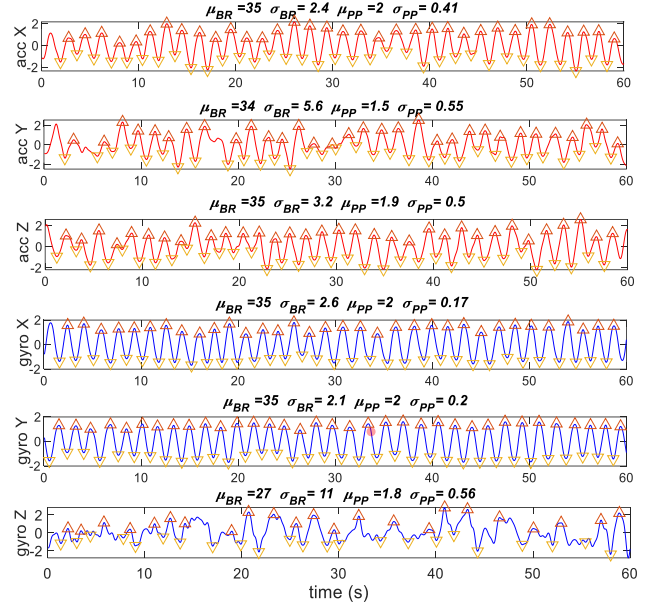

Fig. 6. An example of channel selection for the accelerometer. The optimal channel is 'gyro Y'.

Signal quality cannot be guaranteed during the entire course of overnight recording because patients may have random motion lying on the bed or leave the bed for restroom visits. Various factors such as ambient movement might bring about noises to cause SNR degradation. The position of the wearable sensor to the patient clothing might sometimes move during long or deep breathing, and brought further noise to the signal. Therefore, we opted to remove the epochs with low SNR by pre-determined thresholds. We selected the threshold to be mean of all covariation features to be smaller than 0.4. Table V provides the selection ratio for every dataset and the final cases we have collected after all the signal processing procedures. The datasets from Exp 1 and Exp 2 have higher quality because of the better controlled lab environment during data collection.

TABLE V. STATISTIC COMPARISON OF COVID AND DIFFERENT DATASETS

| | COVID Acc. | Norm. NCS Exp 1 | Exer. NCS Exp 1 | Norm. Acc. Exp 1 | Exer. Acc. Exp 1 | Norm. NCS Exp 2 | Exer. NCS Exp 2 |
|---|---|---|---|---|---|---|---|
| Cases | 10131 | 1049 | 188 | 918 | 231 | 256 | 240 |
| Ratio (%) | 30.2 | 74.0 | 77.7 | 64.7 | 95.5 | 100 | 100 |

IV. RESULTS

*A. Demographic information*

TABLE VI. DEMOGRAPHICS IN HUMAN STUDY

| Datasets | Gender | Number | BMI ($\mu \pm \sigma$) | Age ($\mu \pm \sigma$) |
|---|---|---|---|---|
| COVID | Male | 8 | 30.1 ± 7.3 | - |
| | Female | 4 | 27.8 ± 6.3 | - |
| Exp 1 | Male | 7 | 22.6 ± 2.5 | 29 ± 12 |
| | Female | 6 | 21.4 ± 3.7 | 21 ± 2 |
| Exp 2 | Male | 14 | 23.3 ± 2.5 | 28 ± 9 |
| | Female | 18 | 20 ± 1.3 | 24 ± 2 |

This study involved three distinct datasets: COVID patients, and two dyspnea human studies on healthy subjects. The study population information of each dataset is shown in Table VI. Age information was not gathered for the COVID dataset.

*B. Feature Analysis and Comparison*

After acquiring all datasets, we first evaluated the similarity of respiratory features between COVID patients and healthy subjects where dyspnea was physiologically induced by exercise. The changes of breathing features in exertion-induced dyspnea and acute short-term dyspnea from pulmonary disorders can have correlation with important implications. We first examined a few representative respiratory features and presented the scatter plots in Fig. 7 from 3 datasets: 1) COVID patients (accelerometer); 2) Healthy subjects during normal breathing in Exp. 1 (NCS); 3. Healthy subjects breathing after exertion in Exp. 1 (NCS). Respiratory features collected from NCS and accelerometer in our human study have high similarity, so the difference using two different sources were mainly determined by SNR considerations. We also presented similar scatter plots using the accelerometer in Exp1 and NCS in Exp 2 in supplementary Figs. 1 and 2. In Fig. 7(a), the X and Y axes represented respiratory features $\Re_{BR}$ and $\Re_{PP}$, while in 6(b) represented $\varsigma_{IBI}$ and $\Re_{IBI}$. To better compare the feature distribution for different datasets, we used the Gaussian kernel smoothing function to estimate the returned probability density in top and right lines. The dataset from healthy subjects during normal breathing had a much broader range of distribution compared with the other two datasets. For a better visual demonstration, we set the X-Y limits to only show all of points from dataset 1 and 3, and some points from dataset 2 were out of range. We can observe that COVID patients had a higher similarity of breathing features to healthy subjects after exertion. In Fig. 7(a), for $\Re_{BR}$ and $\Re_{PP}$, and in Fig. 7(b), for $\Re_{IBI}$, both COVID patients and healthy exertional breathing had higher values closer to 1, indicating higher autocorrelation of neighboring breathing cycles. In Fig. 6(b), for $\varsigma_{IBI}$, COVID patients and healthy exertional breathing were concentrated on smaller successive differences, while healthy normal breathing had broader spread to higher variation.

For a more comprehensive comparison of similarity in different datasets, we calculated the Kullback–Leibler (KL) divergence between the dataset of COVID patients and the other datasets in Table VII. The KL divergence, also called the relative entropy, is a type of statistical distance between two probability distributions [31]. We first transformed our discrete datasets to smoothed continuous Gaussian distributions just like the top and right lines in Fig. 7, where the KL divergence represents a natural dissimilarity by the mixture of Gaussians. In Table VII, we presented 8 representative respiratory feature statistics. In both Exp 1 and Exp 2 with healthy subjects, the normal breathing features had larger KL divergence to those of COVID patients than the exertional breathing features. In Table VIII, we also examined the dissimilarity between NCS and accelerometer in the same experiment of Exp 1. The small KL divergence between the two measurements showed the similarity and interchangeability for respiration measurements. Therefore, in the following sections, we will mainly use NCS datasets in Exp 1 for calculation.

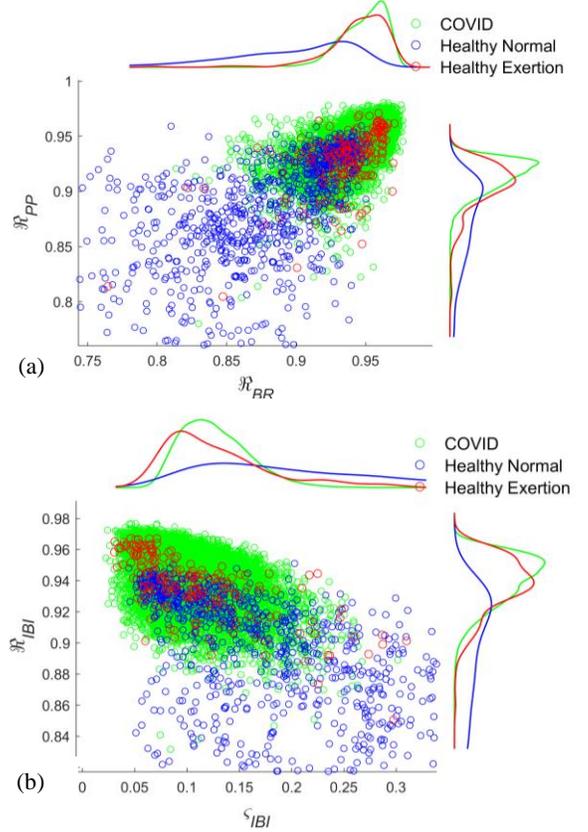

Fig. 7. Scatter plots of chosen respiratory features from COVID and human study datasets. Top and right lines are smoothed continuous distribution by Gaussian kernels. (a): $\Re_{BR}$ and $\Re_{PP}$; (b): $\varsigma_{IBI}$ and $\Re_{IBI}$.

TABLE VII. KL-DIVERGENCE OF COVID TO OTHER DATASETS

| | Norm. NCS Exp 1 | Exer. NCS Exp 1 | Norm. ACC Exp 1 | Exer. ACC Exp 1 | Norm. NCS Exp 2 | Exer. NCS Exp 2 |
|---|---|---|---|---|---|---|
| $\mu_{BR}$ | 2.14 | 0.17 | 2.62 | 0.16 | 1.44 | 0.30 |
| $\sigma_{BR}$ | 1.71 | 0.69 | 0.68 | 0.72 | 0.49 | 0.91 |
| $CoV_{BR}$ | 3.91 | 0.76 | 2.75 | 0.65 | 1.17 | 1.03 |
| $CoV_{IBI}$ | 4.79 | 0.99 | 3.37 | 1.08 | 1.66 | 1.33 |
| $\Re_{BR}$ | 3.42 | 0.16 | 2.97 | 0.16 | 1.52 | 0.50 |
| $\Re_{PP}$ | 2.96 | 0.21 | 2.34 | 0.08 | 0.98 | 0.21 |
| $\varsigma_{IBI}$ | 3.87 | 0.44 | 2.61 | 0.68 | 1.32 | 0.73 |
| $\varsigma_{IER}$ | 1.82 | 0.26 | 1.34 | 0.16 | 0.43 | 0.17 |
| Avg | 3.08 | 0.46 | 2.34 | 0.46 | 1.13 | 0.65 |

TABLE VIII. KL-DIVERGENCE OF NCS AND ACCELEROMETERS IN EXP1.

| | $\mu_{BR}$ | $\sigma_{BR}$ | $CoV_{BR}$ | $CoV_{IBI}$ | |
|---|---|---|---|---|---|
| Norm. | 0.05 | 0.09 | 0.12 | 0.13 | |
| Exer. | 0.01 | 0.13 | 0.11 | 0.06 | |
| | $\Re_{BR}$ | $\Re_{PP}$ | $\varsigma_{IBI}$ | $\varsigma_{IER}$ | Avg. |
| Norm. | 0.04 | 0.05 | 0.12 | 0.21 | 0.10 |
| Exer. | 0.08 | 0.04 | 0.22 | 0.08 | 0.09 |

## C. Dyspnea Classification Model

After comparing individual respiratory features, we can find high similarity between pneumonia-induced dyspnea in COVID patients and exertional dyspnea in healthy subjects. We now adopted our previous dyspnea model derived from Exp 2 [24] on 32 healthy subjects as the training model to evaluate the COVID patients and the healthy subjects with the same sensor setup in Exp 1. By utilizing the ML algorithm in [24], we can evaluate the dyspnea score from all respiratory features as a whole and validate the feasibility of our objective dyspnea reporting system in the clinical setting. As it is impractical to ask patients constantly to self report their dyspnea scores, this objective dyspnea report can be of high value to provide a gauge of dyspnea in COVID patients continuously, especially during inconvenient periods such as patients sleeping or going through treatment.

The first model we built was the binary classification model, namely normal = 0 and dyspnea = 1. For the training dataset we adopted from our previous dyspnea study on healthy subjects, all normal breathing epochs were labelled as normal = 0, while all exertional breathing epochs as dyspnea = 1. By training on the dataset to build a dyspnea classifier, we can output the dyspnea classification results for all COVID datasets and the reference cases from Exp 1. We utilized k-nearest neighbor classifier as the model here with neighbor number $k = 40$. Before feeding the dataset into the model, all features were preprocessed with a standard scaler to regularize features by removing the mean and scaling to unit variance. As Table IX showed, almost all of COVID patients' cases were classified into dyspnea, while 73.6% of exertional breathing cases of healthy subjects were classified into dyspnea. In comparison, only 4.24% of the normal breathing cases in healthy subjects were classified into dyspnea. The dyspnea classification results for COVID patients further corroborated our hypothesis that COVID dyspnea had high similarity to exertional dyspnea in healthy subjects. We also presented similar results using accelerometer in Exp 1 in supplementary Fig. 3.

TABLE IX. CLASSIFICATION RESULTS OF DYSPNEA FOR COVID PATIENTS AND HEALTHY SUBJECTS (IN EXP 1).

|  | COVID | Healthy Normal | Healthy Exertion |
| --- | --- | --- | --- |
| Percentage of Dyspnea | 98.05 % | 4.24 % | 73.63 % |

## D. Dyspnea Scoring Model

In this section, we built a regressor model for objective dyspnea scoring in the Borg scale (0 - 10) $D_{obj}$ for COVID patients. In our previous work, we built a similar dyspnea scoring system and achieved high accuracy for generating $D_{obj}$ for exertional dyspnea on healthy subjects in comparison with self report. In this work, we used our previous dyspnea study as the training model to build the scoring system, and treated the COVID patients as testing cases. We implemented the k-nearest neighbor regressor as the main model here. Since we had overnight recording for COVID patients, we first reported the epoch-wise dyspnea scores for all datasets, and then we averaged all dyspnea scores as the final score for every patient.

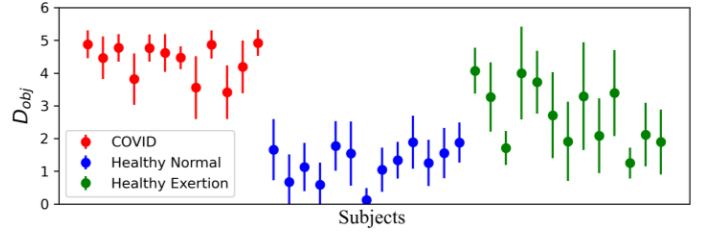

Fig. 8. Dyspnea scoring results for COVID patients and healthy subjects (Exp 1). The average of $D_{obj}$: COVID = 4.39; Healthy Normal = 1.26; Healthy Exertion = 2.72.

TABLE X. T-TEST STATISTICS FOR DYSPNEA SCORING RESULTS

|  | COVID vs. Healthy Normal | COVID vs. Healthy Exertion | Healthy Normal vs. Healthy Exertion |
| --- | --- | --- | --- |
| T-statistic | 14.60 | 5.47 | –4.82 |
| P-value | $4.61 \times 10^{-13}$ | $2.75 \times 10^{-5}$ | $1.1 \times 10^{-4}$ |

Fig. 8 presents the results for dyspnea scoring of COVID patients using exertion induced dyspnea on healthy subjects in Exp 1. Results for benchmark Exp 2 on healthy subjects are also shown. For average $D_{obj}$ reported from different datasets, COVID = 4.39; Healthy Normal = 1.26; Healthy Exertion = 2.72. As observed from Fig. 8, $D_{obj}$ for COVID patients were more concentrated around 4-5, while normal breathing for healthy participants were mostly below 2. Exertional breathing for healthy participants had more participants with higher dyspnea score, but subject variation was also evident, possibly because different subjects had variation in physical conditioning after the same cardio exercise. $D_{obj}$ for COVID patients was less dispersed probably due to the more uniform manifestations of the underlying pulmonary disease. We also presented similar results using the accelerometer in Exp 1, as shown in supplementary Fig. 4.

We further preformed T-tests for dyspnea scoring results on different datasets as shown in Table X. The calculated T-statistic is positive when the sample mean of the first dataset is greater than the second dataset, and negative otherwise. As the T-statistic showed, the dyspnea scores for COVID patients were distinctively higher than normal breathing in healthy subjects, while differences with exertion breathing were smaller. The very small p-value between COVID patients and normal breathing indicated that they had distinctively different distributions for dyspnea scores. For the dataset of healthy exertion, the p-values to COVID and healthy normal were also sufficiently small to suggest high distinguishability among the datasets.

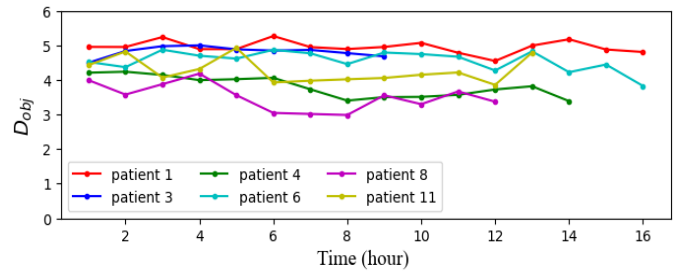

Fig. 9. An example of continuous monitoring of objective dyspnea scores for COVID patients for every hour.

Apart from reporting average $D_{obj}$ to give a general objective evaluation, our system can also output continuous real-time $D_{obj}$ to give an indication to the infection progression. COVID and other important pulmonary diseases like COPD are frequently accompanied by dyspnea sensation from reduced lung function. It is thus critical to continuously monitor patients because the dyspnea often develops insidiously over a period of time. Frequent self report is inconvenient and less accurate for long-term tracing of the symptoms. Fig. 9 shows an example of continuous monitoring of dyspnea score for 6 COVID patients. In the whole recoding of 12-16 hours, $D_{obj}$ was reported every hour, to align with the clinical data recorded every hour.

## V. CONCLUSION

Dyspnea is a key symptom for patients with COVID-19 and many other respiratory disorders. Existing clinical evaluation of dyspnea currently depends on self-report, which is subjective and challenging for continuous monitoring. In this paper, we used an innovative approach to continuously monitor respiratory features using wireless and wearable respiratory sensors to develop an objective dyspnea scoring system derived from exertion routines on healthy subjects. We then tested this model without further learning on COVID patients and control subjects under the same sensor setup. We found high similarity between pneumonia-induced dyspnea of COVID patients and physiologically induced dyspnea on healthy subjects, suggesting that changes of respiratory features from physical exertions could be representative of the dyspnea found in pulmonary disorders. We also demonstrated the unique capability to continuously report objective dyspnea scores during 16 hours for COVID patients. Our system can be a promising tool for diagnosis and prognosis of COVID, offering warning of possible worsening dyspnea and respiratory function, as well as the degree of recovery. This work validates the feasibility of our objective dyspnea scoring for clinical dyspnea assessment, and can be applied to symptomatic evaluation of dyspnea in patients with similar conditions including asthma, pneumonia [32], and COPD [33].